\documentclass[twocolumn,showpacs,aps,prd,preprintnumbers,amsmath,amssymb,superscriptaddress]{revtex4-1}
\usepackage[breaklinks=true]{hyperref}
\usepackage[latin1]{inputenc}
\usepackage[dvips]{graphicx,epsfig,color}
\usepackage{enumerate}
\usepackage{bm}
\usepackage{dcolumn}
\usepackage{url}

\newcolumntype{.}{D{.}{.}{1}}

\newcommand{\ttH}{\ensuremath{{t\overline{t}H}}}

\newcommand{\simgt}{\hbox{ \raise3pt\hbox to 0pt{$>$}\raise-3pt\hbox{$\sim$} }}
\newcommand{\simlt}{\hbox{ \raise3pt\hbox to 0pt{$<$}\raise-3pt\hbox{$\sim$} }}

\begin{document}
\title{
  Measuring the top Yukawa coupling at the ILC at $\bm{\sqrt{\bm{s}}} =$ 500 GeV
}

\author{Ryo~Yonamine}
\affiliation{Department of Particle and Nuclear Physics,
  The Graduate University for Advanced Studies (Sokendai),
  Tsukuba 305-0801, Japan }
\author{Katsumasa~Ikematsu}
\affiliation{Department of Physics, Universit\"at Siegen, D-57068 Siegen, Germany}
\author{Tomohiko~Tanabe}
\affiliation{International Center for Elementary Particle Physics, The University of Tokyo,
  Tokyo 113-0033, Japan}
\author{Keisuke~Fujii}
\author{Yuichiro~Kiyo}
\affiliation{High Energy Accelerator Research Organization (KEK),
  Tsukuba 305-0801, Japan}
\author{Yukinari~Sumino}
\affiliation{Department of Physics, Tohoku University,
  Sendai 980-8578, Japan}
\author{Hiroshi~Yokoya}
\affiliation{Theory Unit, Physics Department, CERN, CH-1211 Geneva, Switzerland}
\affiliation{Department of Physics and National Center for Theoretical Sciences, National Taiwan University, Taipei 10617, Taiwan}

\begin{abstract}
We report on the feasibility of the direct measurement of the top Yukawa coupling $g_t$
at the International Linear Collider (ILC) during its first phase of operation
with a center-of-mass energy of 500~GeV.
The signal and background models incorporate the
non-relativistic QCD corrections which enhance
the production cross section near the $t\overline{t}$ threshold.
The $e^+e^-\to t\overline{t}H$ signal is reconstructed in
the 6-jet + lepton and the 8-jet modes.
The results from the two channels are combined.
The background processes considered are
$e^+e^-\to t\overline{b}W^-/\overline{t}bW^+$
(which includes $e^+e^-\to t\overline{t}$),
$e^+e^-\to t\overline{t}Z$, and
$e^+e^-\to t\overline{t}g^*\to t\overline{t}b\overline{b}$.
We use a realistic fast Monte-Carlo detector simulation.
Signal events are selected using
event shape variables, through jet clustering,
and by identifying heavy flavor jets.
Assuming a Higgs mass of 120~GeV,
polarized electron and positron beams with $(P_{e^-},P_{e^+})=(-0.8,+0.3)$,
and an integrated luminosity of 1~ab$^{-1}$,
we estimate that the $e^+e^-\to\ttH$ events can be
seen with a statistical significance of $5.2\,\sigma$,
corresponding to the relative
top Yukawa coupling measurement accuracy of
$\left|\Delta g_t/g_t\right|=10\%$.
\end{abstract}

\pacs{
  13.66.Jn, 
  14.65.Ha, 
  14.80.Bn, 
}

\maketitle

\section{Introduction}

The standard model (SM) of elementary particle physics stands on two pillars.
The first pillar is the gauge principle, which has been verified
by precision electroweak measurements.
The second pillar consists of the electroweak symmetry breaking
which is yet to be tested by experiment.
The discovery of the Higgs boson will be of particular importance 
in explaining the mass generation mechanism.
Within the SM, the Yukawa interaction of the top
quark and the Higgs boson generates the mass term
which breaks the electroweak gauge symmetry.
The measurement of the strength of the top Yukawa coupling $g_t$ can shed light
on the mechanism behind the generation of the top quark mass.

The top-quark Yukawa interaction could be measured indirectly
using the production mechanism of the Higgs boson
through the top quark loop at the LHC experiments.
The indirect measurement unfortunately cannot give a full description
of the top-quark Yukawa interaction,
for suppose we observe an anomaly in the Higgs production cross section;
it would be difficult to distinguish whether this effect is due to an anomaly in the
top Yukawa interaction itself,
or because there are contributions from unknown
particles propagating in the loop connecting the initial state and the Higgs boson.
In order to distinguish these two effects,
it would be highly desirable to measure the top Yukawa interaction directly.
At the LHC, the direct production process $gg\to t\overline{t}H$
in the $H\to b\overline{b}$ channel is marred by
jet combinatorial background~\cite{Benedetti:2007sn}.
While the $H\to\gamma\gamma$ or $H\to\tau^+\tau^-$ channels
are expected to yield cleaner signals~\cite{Gross:1900zz},
which could allow for the discovery of the $gg\to t\overline{t}H$ process,
the uncertainty in the top Yukawa coupling value
would be affected by the potentially large
uncertainties in the Higgs branching fraction measurements.
We show that a future $e^+e^-$ linear collider, such as the International Linear Collider (ILC),
can play a critical role in the determination of the top Yukawa coupling
through the direct measurement of $e^+e^-\to t\overline{t}H$
in the $H\to b\overline{b}$ channel.

Feasibility studies of the top quark Yukawa interaction at a future $e^+e^-$
linear collider have a long history \cite{Hagiwara:1990dw,Djouadi:1992gp}.
A serious feasibility study of a direct measurement of
the top Yukawa coupling using the process $e^+e^-\rightarrow t\overline{t}H$
at the center-of-mass (CM) energy $\sqrt{s}=800$~GeV was
performed in \cite{Juste:1999af,Gay:2006vs}, which
incorporated realistic experimental conditions 
expected at a linear collider experiment.
More recently, there have been increased interests in
how well the top Yukawa coupling can be measured in the
first phase of a linear collider experiment, whose CM energy
reaches up to $\sqrt{s}=500$~GeV.
An analysis for $\sqrt{s}=500$~GeV was carried out
in \cite{Baer:1999ge}.
It was noted that at this energy the bound-state effects
between $t$ and $\overline{t}$ enhance the
$t\overline{t}H$ production cross section significantly,
since the relative momentum of
$t$ and $\overline{t}$ is typically
small~\cite{Dittmaier:1998dz,Dawson:1998ej,Belanger:2003nm,Denner:2003zp,You:2003zq,Farrell:2005fk,Farrell:2006xe}.
A reanalysis was performed in the Snowmass workshop,
incorporating the enhancement effect by
$t\overline{t}$ resonance formation as well as
an enhancement effect that can be obtained 
by polarizing the $e^+e^-$ beams~\cite{Juste:2006sv}.
The conclusion was that the top Yukawa coupling can be
measured to roughly $10$\% accuracy,
including statistical errors only,
with an integrated luminosity of $1$~ab$^{-1}$.

In this paper, we investigate the feasibility of measuring the top Yukawa coupling at 
$\sqrt{s}=500$~GeV using the process $e^+e^-\rightarrow t\overline{t}H$
for the Higgs mass of 120~GeV. 
The new aspects of this study as compared to the previous ones are as follows.
We implement the enhancement factor by $t\overline{t}$ bound-state effects
into the event generator, both for the $t\overline{t}H$ signal and
the $t\overline{t}Z$ background events;
the latter is particularly important since the expected measurement accuracy
of the top Yukawa coupling
is significantly affected by the number of these background events.
In addition, we perform a fairly detailed detector simulation
which takes into account the realistic energy resolution
of the calorimeter components. (See Sec.~\ref{sec:framework}).
 
This paper is organized as follows.
In Sec.~\ref{sec:ttbar-bound-state},
we present our method for including the enhancement by
$t\overline{t}$ bound-state effects.
In Sec.~\ref{sec:signal-background}, the signatures of the $t\overline{t}H$ process
and the possible background processes are outlined.
The analysis framework used for the event generation and the
detector simulations is discussed in Sec.~\ref{sec:framework}.
We discuss the event selection procedure in detail
in Sec.~\ref{sec:6jet} for the 6-jet plus lepton mode analysis,
and in Sec.~\ref{sec:8jet} for the 8-jet mode.
We summarize the accuracy estimate of the top Yukawa coupling measurement
in Sec.~\ref{sec:conclusions}.
The measurement is assumed to be dominated by the statistical uncertainty.

\section{Inclusion of $\bm{t}\overline{\bm{t}}$ bound-state effects}
\label{sec:ttbar-bound-state}

Theoretical analyses of the $t\overline{t}$ bound-state effects
on the cross section for $e^+e^-\rightarrow t\overline{t}H$ with the
next-to-leading logarithmic accuracy are given in \cite{Farrell:2006xe}.
To our knowledge, there have been no
analyses of $t\overline{t}$ bound-state effects for the background process
$e^+e^-\rightarrow t\overline{t}Z$
which are included consistently with the
$e^+e^-\rightarrow t\overline{t}H$ signal.
In our event generator, all the QCD corrections are
incorporated consistently only with the leading-order accuracy.
Hence, for consistency, we incorporate the $t\overline{t}$ bound-state effects
on the signal and background cross sections
with the leading-order accuracy;
we also incorporate some of
the important next-to-leading order corrections in the bound-state effects.

There are a number of (tree-level) Feynman diagrams contributing to
each of the processes
$e^+e^-\rightarrow t\overline{t}H$,
$e^+e^-\rightarrow t\overline{t}Z$ and 
$e^+e^-\rightarrow t\overline{t}g^*$,
where the $t(\overline{t})$ subsequently decays into $bW^+(\overline{b}W^-)$.
Let us denote these amplitudes for the process $i\to f$
as $A_{t\overline{t}}(i\to f)$.
The tree level amplitudes are modified as follows:
\begin{align}
\nonumber
A_{t\overline{t}}(i\to f) &= [A_{t\overline{t}}(i\to f)]_{\mathrm{tree}} \\
&\times \sqrt{ K_{i\to f} } \times F({\hat{s}_{t\overline{t}}},\vec{p};m_t,\Gamma_t,\alpha_s).
\label{defmodfac}
\end{align}
$F$ represents a process-independent enhancement factor that incorporates
$t\overline{t}$ $S$-wave bound-state effects;
$\sqrt{\hat{s}_{t\overline{t}}}$ denotes the CM energy of $t$ and $\overline{t}$
as reconstructed from the final $bW^+\overline{b}W^-$ system;
$\vec{p}$ is the three-momentum of $t$ in the CM frame of $t$ and $\overline{t}$;
$m_t$ and $\Gamma_t$ denote the pole mass and width of the top quark,
respectively.
Close to the threshold of $t\overline{t}$ pair production,
this factor $F$ incorporates the bound-state effects according to the
non-relativistic bound-state theory,
while for higher values of $\sqrt{\hat{s}_{t\overline{t}}}$,
the factor $F$ is smoothly interpolated to unity:
\begin{eqnarray}
F = \left\{
\begin{array}{lll}\displaystyle
\frac{G(E,\vec{p})}{G_0(E,\vec{p})},
&~~~&
E\equiv \sqrt{\hat{s}_{t\overline{t}}} - 2m_t \ll m_t ,
\\
\rule[0mm]{0mm}{7mm}
1, && E \simgt m_t .
\end{array}
\right.
\end{eqnarray}
The non-relativistic Green function is defined by
\begin{eqnarray}
&&
\left[ (E+i\Gamma_t)-
\left\{ - \frac{\nabla^2}{m_t} + V_\mathrm{QCD}(r)\right\}\right]
G(E,\vec{x}) = \delta^3(\vec{x}) ,
\label{defGreenfn}
\\
&&
G(E,\vec{p})=\int d^3\vec{x}\, e^{-i\vec{p}\cdot\vec{x}}G(E,\vec{x}) ,
\label{defGreenfn2}
\end{eqnarray}
where $r=|\vec{x}|$ and
$V_\mathrm{QCD}(r)$
is the next-to-leading order QCD potential \cite{Farrell:2006xe}.
$G_0(E,\vec{p})$ is the non-relativistic Green function of
a free $t\overline{t}$ pair, which is defined via Eqs.~(\ref{defGreenfn})
and (\ref{defGreenfn2})
after setting $V_\mathrm{QCD}(r)$ to zero.
The enhancement factor $F$ is explained in more detail
in Ref.~\cite{Sumino:2010bv}.

In Eq.~(\ref{defmodfac}) $K_{i\to f}$ denotes a
process-dependent hard-vertex correction factor, which arises as
a part of the next-to-leading order corrections.
To a good approximation this factor is
independent of kinematical variables for the
signal process
$e^+e^- \to t\overline{t} H$.
In fact with a choice $K_{i\to f}=0.843$, we reproduce
the $e^+e^-\to t\overline{t}H$ differential
cross section at $\sqrt{s}=500$~GeV shown in Fig.~5(a) of
Ref.~\cite{Farrell:2005fk}.
(The next-to-leading-logarithmic curve in the threshold region
of $t\overline{t}$ and the ${\cal O}(\alpha_s)$
curve at higher $\sqrt{\hat{s}_{t\overline{t}}}$ or lower $E_H$.)
We adopt this value of $K_{i\to f}$ for the signal process.
For the background processes, we choose $K_{i\to f}=1$,
since these factors are unknown and since the deviation of these factors
from unity is part of the next-to-leading
order corrections not fully accounted for in our analysis.

\section{Signal and background processes}
\label{sec:signal-background}
\begin{figure}[hbt!p]
\centering
\includegraphics[width=0.45\linewidth]{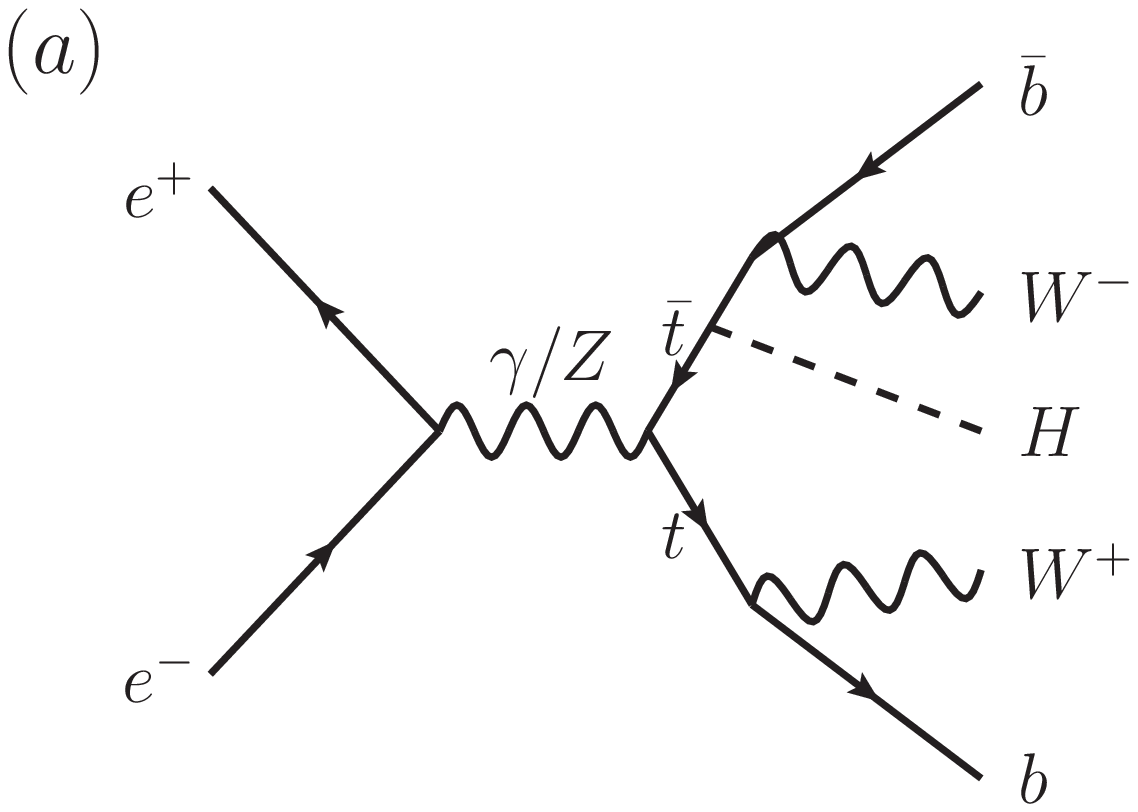}
\includegraphics[width=0.45\linewidth]{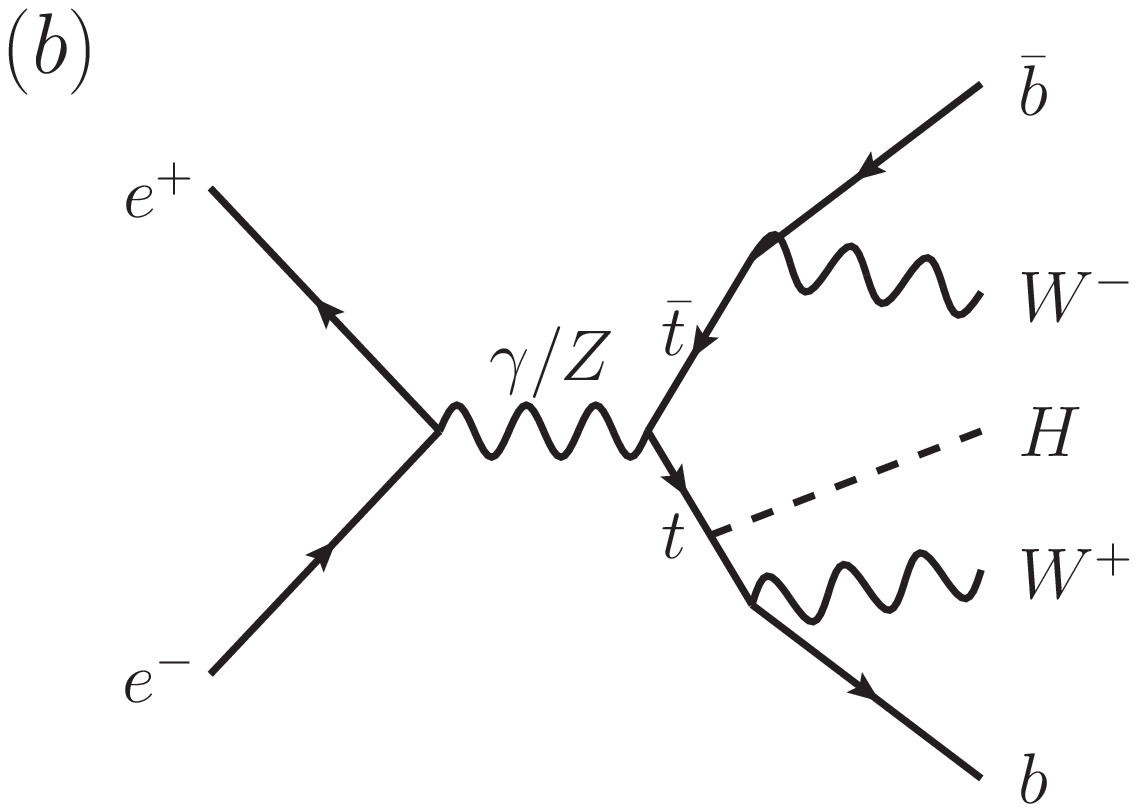}
\includegraphics[width=0.45\linewidth]{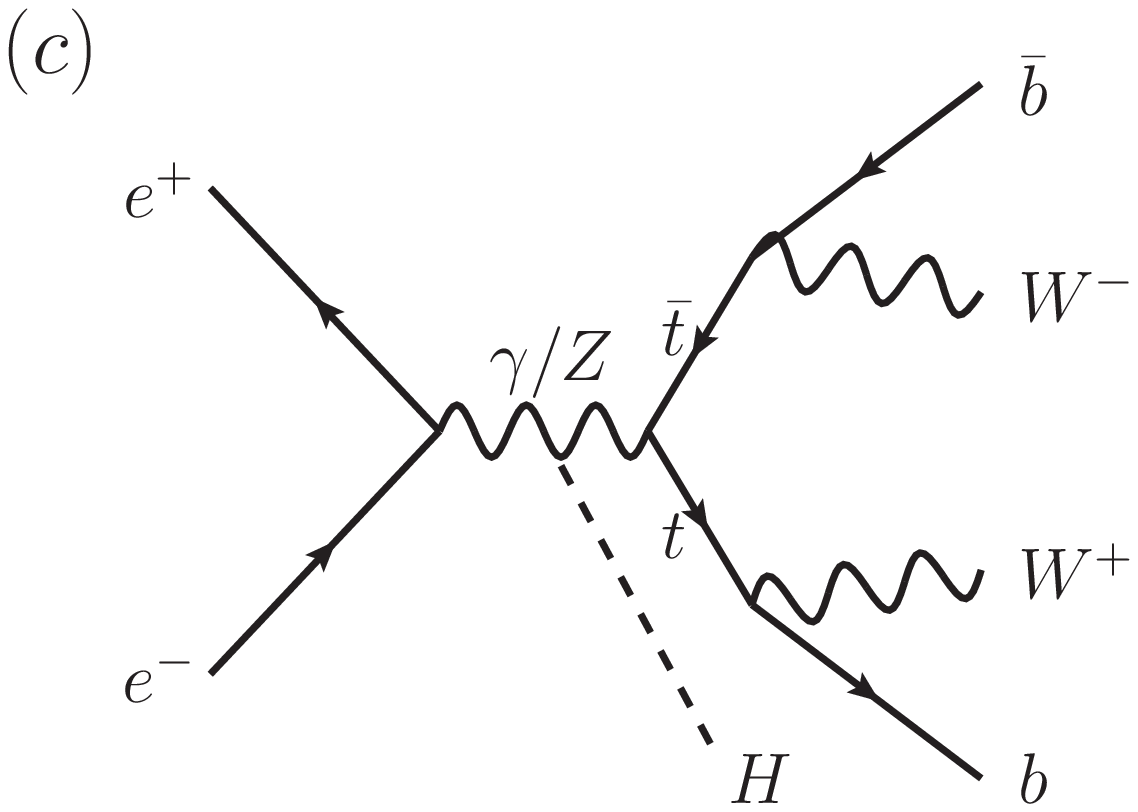}
\caption{ Feynman diagrams for the $e^+e^-\to t\overline{t}H$ process.  }
\label{fig:TTHdiagram}
\end{figure}

\begin{figure}[hbt!p]
\begin{center}
\includegraphics[width=0.9\linewidth]{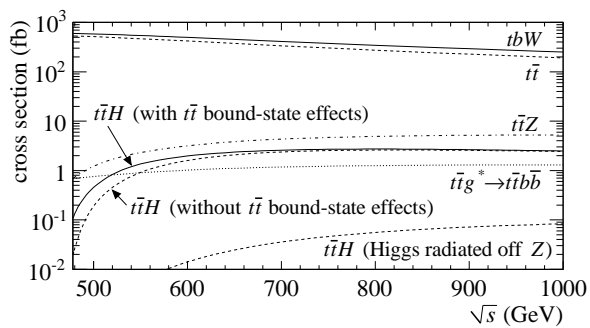}
\caption{
  Production cross section of the
    $e^+e^-\to t\overline{t}H$ signal
    (shown with and without $t\overline{t}$ bound-state effects),
    together with those of the main background processes, 
  $t\overline{t}H$ (Higgs radiated off the $Z$ boson),
  $t\overline{t}Z$,
  $t\overline{t}$,
  $t\overline{b}W^- / \overline{t}bW^+$ (denoted as $tbW$),
  and
  $t\overline{t}g^*\to t\overline{t}b\overline{b}$,
  as a function of the CM energy without beam polarizations.
    The initial state radiation and beamstrahlung effects are included.
}
\label{fig:xsection}
\end{center}  
\end{figure}

The diagrams for $e^+e^-\to t\overline{t}H$ with subsequent top decays
$t\to bW^+~(\overline{t}\to\overline{b}W^-)$
are shown in Fig.~\ref{fig:TTHdiagram}.
The first and second diagrams containing the top Yukawa coupling $g_t$
are the targets of this study.
The contribution to the cross section coming from the third diagram,
where the Higgs radiates off of the intermediate $Z$ boson,
is negligible at $\sqrt{s}=500$~GeV,
as shown in Fig.~\ref{fig:xsection}.
As a result, the number of $\ttH$ events is proportional to $g_t^2$
to a very good approximation,
which enables us to perform a simple analysis by event counting.

In this study, the Higgs boson is reconstructed
in the two $b$-jet mode $H\rightarrow b\overline{b}~(68\%)$.
Our $\ttH$ signal can be classified into three groups,
depending on the decay mode of the $W$ bosons.
Their branching fractions are
\begin{enumerate}[(i)]
  \item 8-jet mode: 45\%,
  \item 6-jet + lepton mode ($e$ or $\mu$): 29\%,
  \item 4-jet + 2-lepton mode ($ee$, $e\mu$, or $\mu\mu$): 5\%,
\end{enumerate}
where we have omitted the contribution of the top decays to tau
($t\to b\tau^+\nu_\tau$ and $\overline{t}\to \overline{b}\tau^- \overline{\nu}_\tau$),
since we only reconstruct electrons and muons from the top in this study.
The 8-jet mode and the 6-jet + lepton mode are chosen for reconstruction.

The following processes are identified as possible background sources
which can mimic the $\ttH$ signatures:
\begin{enumerate}[(i)]
\item $e^+e^-\to t\overline{b}W^-/\overline{t}bW^+\to bW^+\overline{b}W^-$,
\item $e^+e^-\to t\overline{t}Z\to bW^+\overline{b}W^-b\overline{b}$,
\item $e^+e^-\to t\overline{t}g^*\to bW^+\overline{b}W^-b\overline{b}$.
\end{enumerate}
The cross sections for these processes are shown
as a function of $\sqrt{s}$ in Fig.~\ref{fig:xsection}.
We will refer to the
$e^+e^-\to t\overline{b}W^-/\overline{t}bW^+$
process as $e^+e^-\to tbW$.
The $e^+e^-\to tbW$ process includes the
$e^+e^-\to t\overline{t}$.
The $e^+e^-\to tbW$ final state
consists of up to two $b$~jets, as opposed to
four $b$~jets for our $\ttH$ signal.
The $tbW$ channel can be therefore reduced to a small fraction
by identifying the flavor of the $b$~quarks
in the final state ($b$-tagging)
and by counting the number of $b$~jets.
Because of the large $tbW$ cross section,
a significant amount of $tbW$ background remains
even if there is a small rate of event mis-reconstruction,
which occurs equally likely for events in
and away from the top pair resonance,
thus making it important to include
the non-resonant contributions.

In contrast to the $tbW$ process,
the processes $t\overline{t}Z$ and
$t\overline{t}g^*$
can have identical final states as those of the $\ttH$ process
if the $Z$ boson or the hard gluon $g^*$ decays into a $b\overline{b}$ pair.
In this case, the signal extraction will depend strongly on the
resolution of the Higgs mass reconstructed from the two $b$-jets.
The unpolarized cross section for $t\overline{t}Z$
is 1.3~fb, including the $t\overline{t}$ bound-state effects
similar to that expected for the signal process;
without including this correction, the cross section becomes 0.7~fb.
For $t\overline{t}g^*\to t\overline{t}b\overline{b}$,
the unpolarized cross section is 0.7~fb.
We note that there is no $t\overline{t}$ bound-state enhancement
in the $t\overline{t}g^*$ process because
the $t\overline{t}$ system is not a color singlet in this case.
The cross sections at $\sqrt{s}=500$~GeV for our signal and
background processes are summarized in Tab.~\ref{tab:Xsection}.
\begin{table}[hbt!p]
\caption{Cross sections at $\sqrt{s}=500$~GeV for the signal and background processes
  are shown for the different beam polarizations.
  The $e^+e^-\to t\overline{t}H$ and $e^+e^-\to t\overline{t}Z$ processes include the
  $t\overline{t}$ bound-state effects.
  The $t\overline{t}H$, $t\overline{t}Z$, and $t\overline{t}g^*$ processes all decay as
  $bW^+\overline{b}W^-b\overline{b}$
  while the $\overline{t}bW^+/t\overline{b}W^-$ process (denoted as $tbW$) decays as
  $bW^+\overline{b}W^-$.
  The number of events $N$ used in this study is shown for each sample,
  along with its equivalent luminosity $\mathcal{L}$.
}
\label{tab:Xsection}
\centering
\begin{ruledtabular}
\begin{tabular}{lcc.}
Process & \multicolumn{1}{c}{$\sigma$ (fb)} & $N$ & \multicolumn{1}{c}{$\mathcal{L}$ (ab$^{-1}$)} \\
\hline
$e^{-}_{L}e^{+}_{R}\to t\overline{t}H$   &  1.07& $5.00\times10^4$ & 47.8\\
$e^{-}_{L}e^{+}_{R}\to t\overline{t}Z$   &  4.04& $5.00\times10^4$ & 12.4\\
$e^{-}_{L}e^{+}_{R}\to t\overline{t}g^*$ &  1.93& $5.00\times10^4$ & 25.9\\
$e^{-}_{L}e^{+}_{R}\to tbW$              &1633  & $1.00\times10^7$ &  6.1\\
\hline
$e^{-}_{R}e^{+}_{L}\to t\overline{t}H$   &  0.45& $5.00\times10^4$ & 92.6\\
$e^{-}_{R}e^{+}_{L}\to t\overline{t}Z$   &  1.32& $5.00\times10^4$ & 37.8\\
$e^{-}_{R}e^{+}_{L}\to t\overline{t}g^*$ &  0.86& $5.00\times10^4$ & 58.2\\
$e^{-}_{R}e^{+}_{L}\to tbW$              &700   & $1.00\times10^7$ & 14.3\\
\end{tabular}
\end{ruledtabular}
\end{table}
The signal and background samples have been produced
with pure beam polarizations.
Unless otherwise noted, our results weight these samples to match the
beam polarizations of $(P_{e^-},P_{e^+})=(-0.8,+0.3)$~\cite{Brau:2007zza}.

\section{Analysis framework}
\label{sec:framework}
\begin{figure}[hbt!p]
\centering
\includegraphics[width=0.9\linewidth]{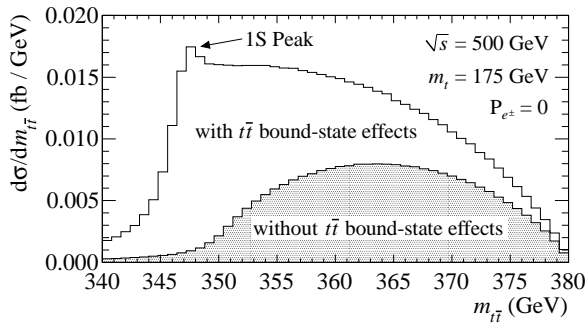}
\caption{
  Differential production cross section as a function of the
  invariant mass of the $t\overline{t}$ system
  for unpolarized beams $(P_{e^-},P_{e^+})=(0.0,0.0)$
  with $\sqrt{s}=500$~GeV.
  The shaded histogram represents the leading-order values.
  The white histogram includes the $t\overline{t}$ bound-state effects.
}
\label{fig:qcdcorrection}
\end{figure}

Signal and background events are generated using
the {\tt physsim}~\cite{physsim} event generator,
based on the full helicity amplitudes including gauge boson decays,
calculated using {\tt HELAS}~\cite{Murayama:1992gi}
and {\tt BASES}~\cite{Kawabata:1985yt},
which properly takes into account
the angular distributions of the decay products. 
For the event generation, the following values are used:
$\alpha(M_{Z}) = 1/128$,
$\sin^2\theta_W=0.230$,
$\alpha_{s} = 0.120$, 
$M_{W}=80.0~\text{GeV}$,
$M_{Z}=91.18~\text{GeV}$,
$M_{t}=175~\text{GeV}$, 
and $M_{H}=120~\text{GeV}$.
The effects of initial state radiation and beamstrahlung are included.
The $t\overline{t}$ bound-state effects
results in a roughly twofold increase in the $\ttH$ signal cross section
at $\sqrt{s}=500\,$GeV,
as shown in Fig.~\ref{fig:qcdcorrection}.
The four-momenta of the final-state quarks and leptons are passed as input to
{\tt Pythia}~6.4~\cite{Sjostrand:2003wg} for parton showering and hadronization. 
The detector response is simulated using
the {\tt QuickSim}~\cite{jsf} fast Monte-Carlo detector simulator.

The detector consists of the beam pipe,
a vertex detector, a drift chamber,
  an electromagnetic calorimeter (ECAL),
  and a hadronic calorimeter (HCAL).
The crossing angle of the beams are also taken into account.
Each hit in the tracking detector is smeared according to the
detector resolution specified in Tab.~\ref{Table:detparam}.
For each charged particle, the parameters describing its helical trajectory
are smeared according to the full covariance matrix of the parameters.
Calorimeters are simulated down to the level of individual cells
with possible overlaps of energy deposits from nearby particles.
Each hit in the calorimeter cell is smeared according to Tab.~\ref{Table:detparam}.
The calorimeter cell hits are clustered
and then matched to the tracks of charged particles.
ECAL clusters which are consistent with a charged track are
subtracted based on the particle flow approach~\cite{Brient:2002gh}.
HCAL clusters whose energy is consistent with charged hadrons are removed,
while for clusters with inconsistent energy matching,
as is the case when neutral hadrons are present,
HCAL energy deposits are statistically subtracted
by an amount weighted by the geometrical overlap
between the charged hadrons and the HCAL clusters.

\begin{table}[hbt!p]
\caption{Detector parameters. $p, p_{T}$ and $E$ are measured in units of GeV.
The angle $\theta$ is measured from the beam axis.}
\label{Table:detparam}
\centering
\begin{ruledtabular}
\begin{tabular}{llc}
Detector	       &Resolution                                                      &Coverage\\
\hline
Vertex detector        & $\sigma_b = 7.0 \oplus (20.0/p\sin^{3/2}\theta)$ $\mu$m			& $|\cos\theta| \le 0.90$	\\
Drift chamber          & $\sigma_{P_{T}}/P_{T} = 1.1 \times 10^{-4}p_{T} \oplus 0.1\%$	& $|\cos\theta| \le 0.95$ \\
ECAL                   &$\sigma_{E}/E = 15\%/\sqrt{E} \oplus	1\%$						& $|\cos\theta| \le 0.90$	\\
HCAL                   &$\sigma_{E}/E = 40\%/\sqrt{E} \oplus	2\%$						& $|\cos\theta| \le 0.90$	\\
\end{tabular}
\end{ruledtabular}
\end{table}

\section{Analysis of the 6-jet + lepton mode}
\label{sec:6jet}
We describe the event selection for the analysis of the
6-jet + lepton mode first, followed by the 8-jet mode.
Similar techniques are used in both modes.
The main differences between the two analyses are
the presence of a lepton in the 6-jet + lepton mode
and the number of jets in the final state.

\subsection{Identification of an isolated lepton}
In the 6-jet + lepton mode,
the lepton from the $W\to\ell\nu$ tends to be energetic
and isolated from the rest of the event.
To identify such a lepton ($e$ or $\mu$),
a cone with a half-opening angle $\theta_\mathrm{cone}$
is constructed around each track (lepton candidate).
The cone energy $E_\mathrm{cone}$ is defined to be the sum of the energy
of all the tracks inside the cone, excluding the lepton candidate.
The value of $\theta_\mathrm{cone}$ which gives
$\cos\theta_\mathrm{cone}=0.98$ is found to be optimal
for our event selection.
\begin{figure}[hbt!p]
\centering
\centering
\includegraphics[width=0.9\linewidth]{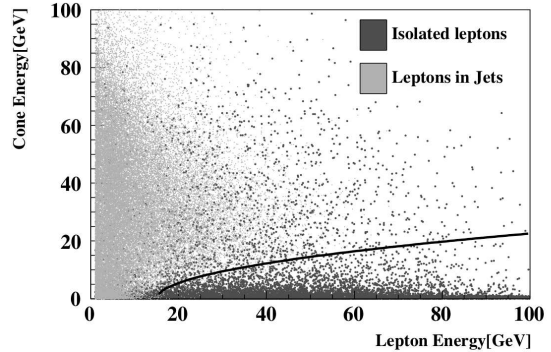}
\caption{Distribution of the cone energy and the lepton energy,
  shown for leptons from $W$ decays (black dots)
  and leptons originating in heavy flavor jets (gray dots).
  The black curve shows the cut boundary for the lepton selection;
  leptons below the curve are identified as isolated leptons.
}
\label{fig:isolepton}
\end{figure}  
Fig.~\ref{fig:isolepton} shows the distribution of the cone energy
versus the lepton candidate energy.
The energetic isolated leptons from the $W$ decay have a
high lepton energy and a low cone energy,
thus populating the lower right region, shown as black dots
in Fig.~\ref{fig:isolepton}.
Leptons from heavy flavor jets are likely to be less energetic
and have a higher cone energy, shown as gray dots in Fig.~\ref{fig:isolepton}.
The selection of isolated leptons is performed by
applying a cut on the cone energy
which varies as the lepton energy and is given by the equation
$E_{\mathrm{cone}} < \sqrt{ 6(E_\ell-15) }$
where $E_{\mathrm{cone}}$ and $E_\ell$ are given in units of GeV.
For the 6-jet + lepton event selection, we require one and only one isolated lepton.

\subsection{Event shape}
To exploit the differences in the event topology between
$e^+e^-\to\ttH$ and $e^+e^-\to tbW$ events,
we use the \emph{thrust} $T$, defined as~\cite{Brandt:1964sa,Farhi:1977sg}
\begin{equation}
T=\max_{|\hat{n}|=1}\frac{\sum_i|\hat{n}\cdot\vec{p}_i|}{\sum_i|\vec{p}_i|},
\end{equation}
where the index $i$ runs over each reconstructed particle
with $\vec{p}_i$ corresponding to its momentum,
and $T$ is to be maximized with respect to the unit vector $\hat{n}$
corresponding to the axis in which the overall event shape is stretched.
The thrust $T$ tends to unity for 2-jet-like events
while it tends to $1/2$ for isotropic events.
Because the $t\overline{t}$ events tend to have fewer jets,
with a higher average jet energy,
$T$ tends to be higher for $tbW$ events compared to $\ttH$ events.
The requirement of $T<0.77$ is found to be optimal in the 6-jet + lepton analysis.

\subsection{Jet clustering}
We employ the Durham jet clustering algorithm~\cite{Catani:1991hj}
to separate the event into 6 jets, after taking out the isolated lepton.
In the Durham algorithm, each particle is regarded as a jet on its own to begin with;
a jet pair $i$ and $j$ gets combined if the pair has the lowest
$Y_{ij}$ value which is defined as
\begin{eqnarray}
Y_{ij} = \frac{\min\{E_i^2,E_j^2\}(1-\cos\theta_{ij})}{E_\mathrm{cm}^2},
\end{eqnarray}
where $\theta_{ij}$ is the angle between the momentum vectors of the two jets,
      and $E_\mathrm{cm}=\sqrt{s}$.
The jet clustering is allowed to continue until there are two jets remaining,
with the value of $Y_{ij}$ being recorded at each transition from
$n$ jets to $n-1$ jets, which we call $Y_{n\to n-1}$.
It is found that $Y_{5\to4}$ is useful for
discriminating $\ttH$ 6-jet + lepton events from $tbW$ events.
This is due to the fact that, after identifying the isolated lepton $\ell$
from the semileptonic decay $t\to bW\to b\ell\nu$,
$tbW$ events cannot have more than four jets without a gluon emission.
Taking all the backgrounds into account,
we require $Y_{5\to4}>0.005$.
The jet configuration for $n=6$ is used
for the rest of the analysis of the 6-jet + lepton mode.

\subsection{Identification of heavy flavor jets}
The identification of jet flavor is critical for the suppression of
$tbW$ background due to the differences in the number of
$b$-jets in the final state between
$tbW$ events, which produce two $b$-jets,
and our $\ttH$ signal, which results in four $b$-jets.
The jet flavor is identified by
looking at the number of secondary tracks belonging to the jet.
We count the number of tracks
with an impact parameter significance (in three dimensions)
greater than a certain threshold value $Q$;
if the count is equal to or greater than a certain number $N_Q$,
the jet is identified as a $b$-jet.
The two numbers $(Q,N_Q)$ are optimized in our
$b$-tagging selection.

In the 6-jet + lepton analysis, we define two criteria for
the identification of $b$-jets.
We define
\emph{loose} $b$-jets as jets passing the
$b$-tagging requirement of $(Q,N_Q)=(2.0,2)$;
\emph{tight} $b$-jets are defined by
those passing the $b$-tagging requirement of $(Q,N_Q)=(2.5,4)$.
Note that the set of tight $b$-jets 
is a subset of loose $b$-jets.
We require at least four loose $b$-jets in the event;
two out of the four are also required to pass the tight $b$-jet criteria.

The $b$-tagging efficiency is estimated using
a sample of $Z\to q\overline{q}$ events at $\sqrt{s}=91.2$~GeV
and is found to be 81\% (loose) and 47\% (tight).
The rate of incorrectly identifying a jet originating from
a lighter quark as a fake $b$-jet is
estimated to be 40\% (loose) and 3.2\% (tight) in a sample of $c$-jets,
and 0.5\% (loose) and 0.08\% (tight) for $s$, $d$, and $u$-jet samples.
In a multi-jet environment, the $b$-tagging efficiencies
decrease due to overlapping jets.
For 6-jet + lepton and 8-jet events, this effect typically
reduces the $b$-tagging efficiencies by roughly 10\%.
In principle, given the fact that the
$b\to c$ mis-tagging rate is rather large,
final states involving charm quark jets
such as the process $e^+e^-\to t\overline{t}g^*\to bW^+\overline{b}W^-c\overline{c}$
can be a source of background.
This effect is not included in our analysis.
However, we expect that the use of a more sophisticated $b$-tagging method
will reduce the charm contamination in the selected $b$-jets~\cite{Bailey:2009ui}.

\subsection{Top and Higgs reconstruction}
The Higgs candidate ($H\to b\overline{b}$) is formed by
requiring one tight $b$-jet and one loose $b$-jet.
The hadronic top candidate is formed by combining
three jets, one of which must be (at least) a loose $b$-jet.
Because there are multiple possible ways to combine the six jets in this way,
we define the quantity $\chi^2$ as
\begin{equation}
\chi^2 =
\left(\frac{m_{j_1j_2}-M_H}{\sigma_H}\right)^2
+\left(\frac{m_{j_3j_4j_5}-M_t}{\sigma_t}\right)^2
+\left(\frac{m_{j_3j_4}-M_W}{\sigma_W}\right)^2,
\label{eq:chi2-6j}
\end{equation}
and choose the jet combination which minimizes the $\chi^2$ value.
Here, $m_{jj}$ $(m_{jjj})$ is the invariant mass of the two-jet (three-jet) system;
the two jets $j_1$ and $j_2$ are used to form the Higgs candidate,
while $j_3$, $j_4$, and $j_5$ are the three jets
used to reconstruct the top candidate which decays hadronically.
The masses $M_t$, $M_W$, and $M_H$ are taken to be the same values
used in the event generation.
The widths $\sigma_t$, $\sigma_W$, and $\sigma_H$ correspond to the mass resolutions
in the case of perfect jet clustering and jet combinations.
These values are determined to be
$\sigma_t=14.3$~GeV, $\sigma_W=9.3$~GeV, and $\sigma_H=17.7$~GeV
by combining the reconstructed four-momenta
of final particles
in the $e^+e^-\to t\overline{t}H$ sample
using Monte-Carlo information.

Final cuts are applied to the resulting invariant mass distributions.
For the 6-jet + lepton mode, we require the top mass to be in the range of
$140<m_t<205$~GeV and the Higgs mass to be in the range of
$95<m_H<150$~GeV,
where the range has been optimized in steps of 5~GeV.

\subsection{Results}
\begin{figure*}[hbt!p]
\centering
\includegraphics[width=0.45\linewidth]{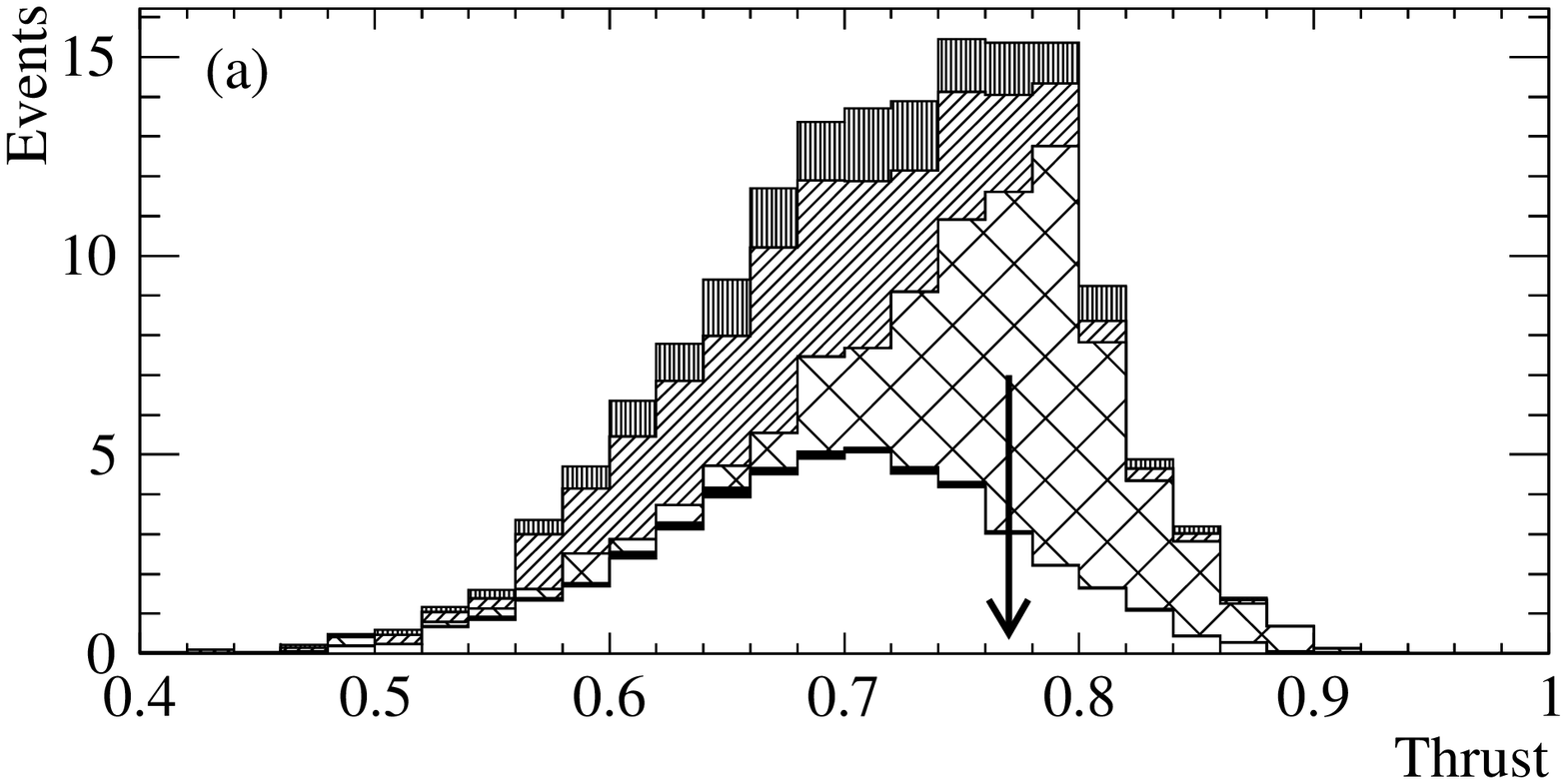}
\includegraphics[width=0.45\linewidth]{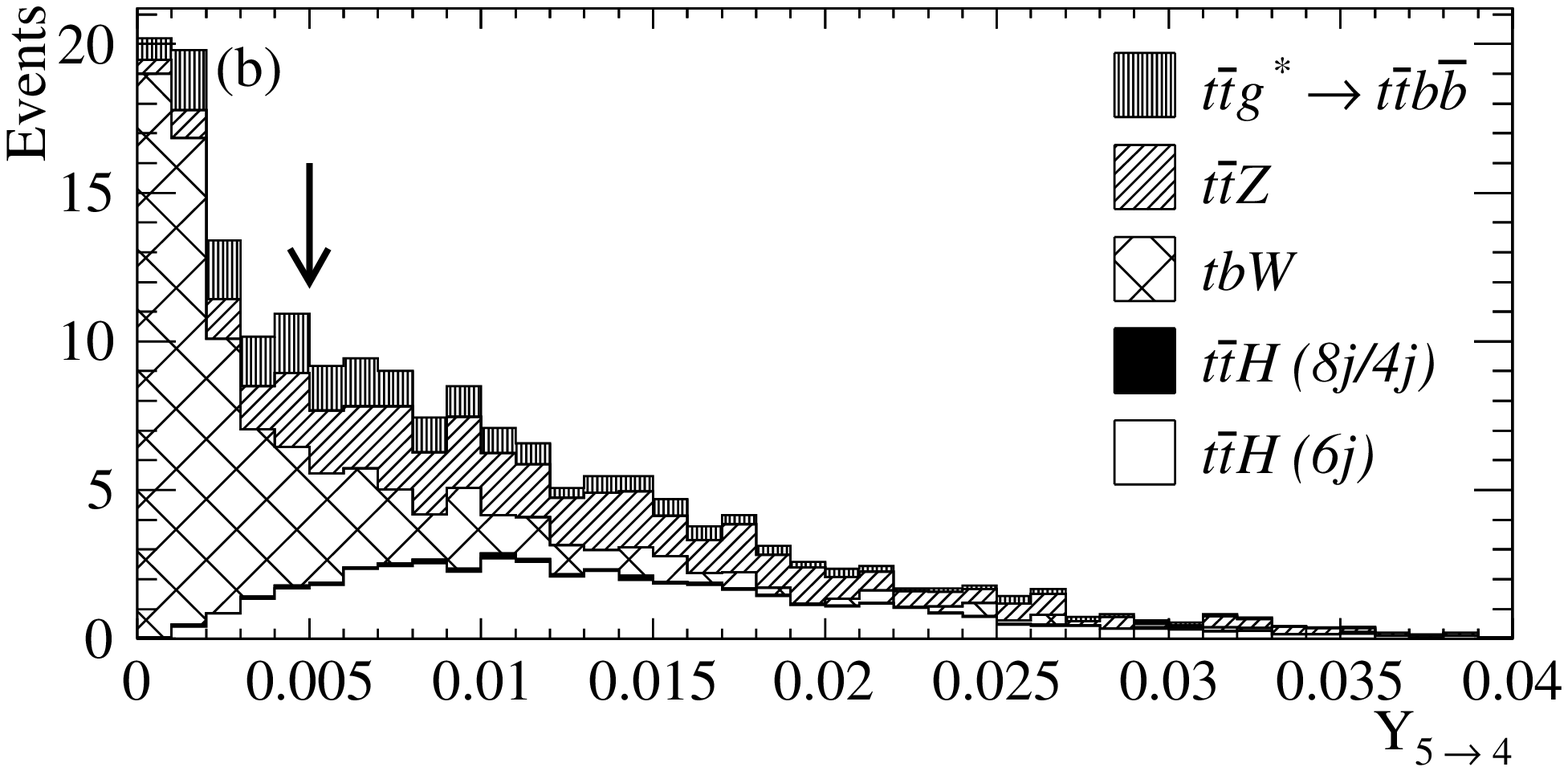}
\includegraphics[width=0.45\linewidth]{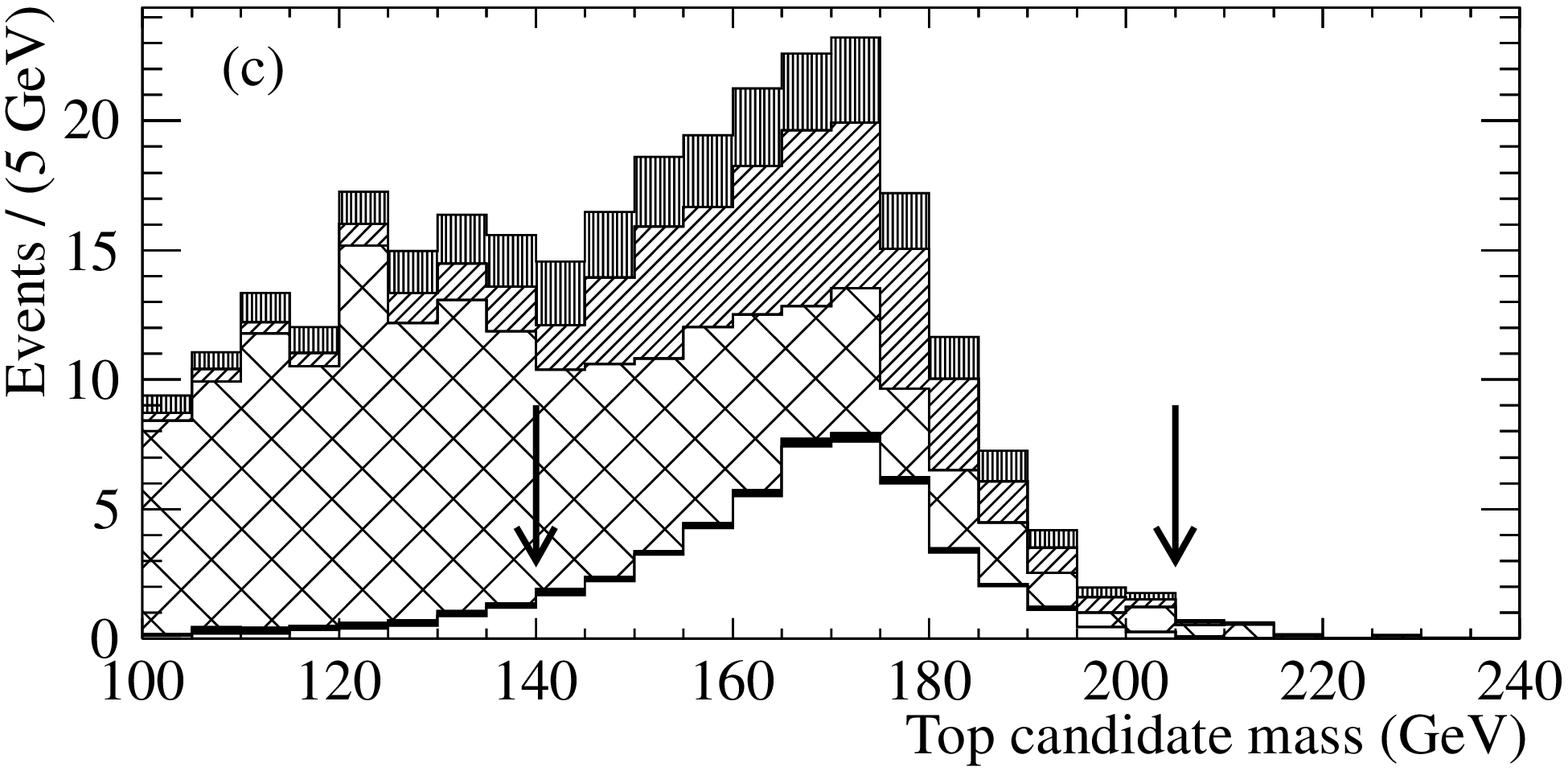}
\includegraphics[width=0.45\linewidth]{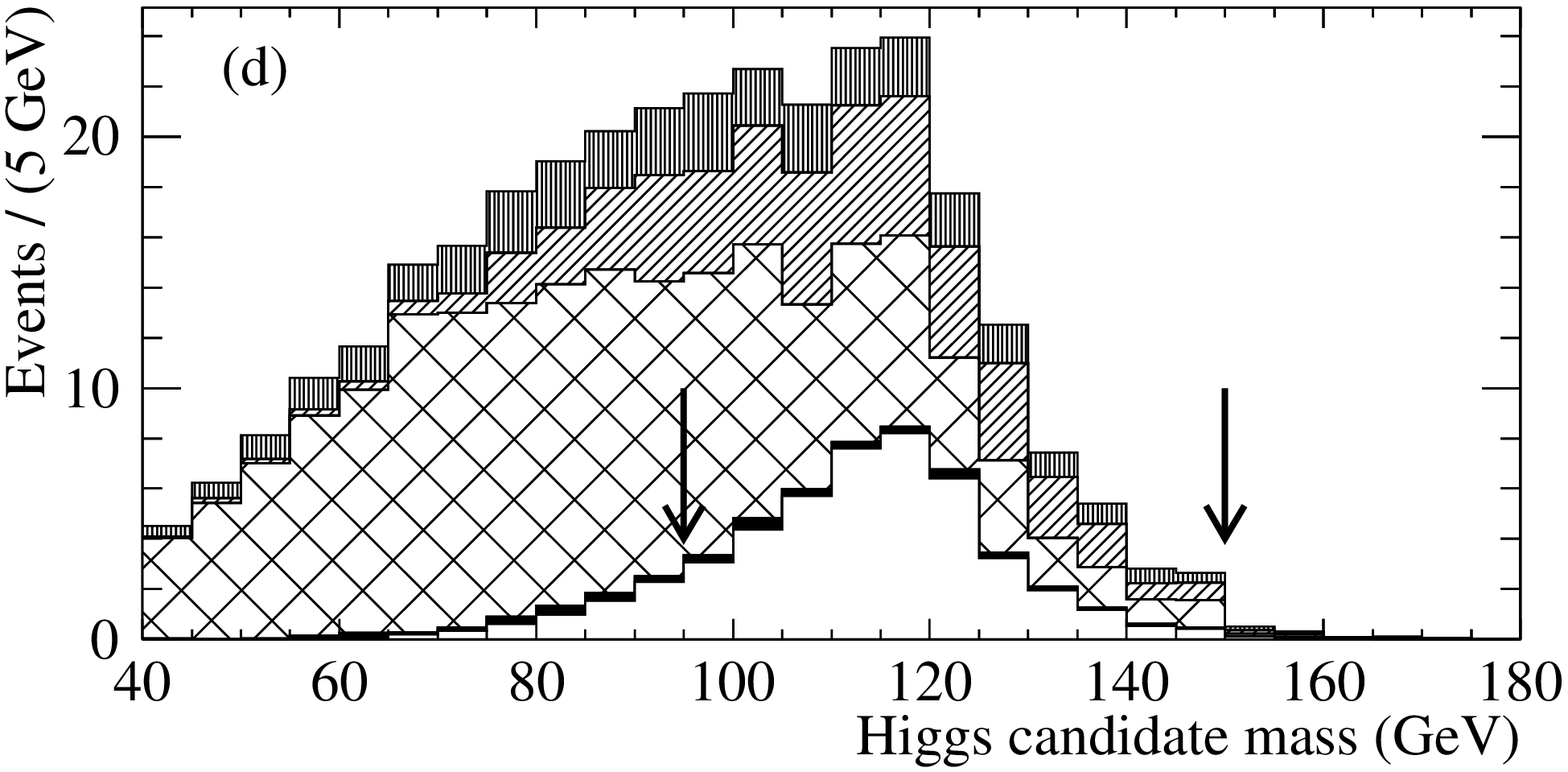}
\caption{ 
The distributions of the cut variables in the 6-jet + lepton analysis are shown:
(a) thrust, (b) $Y_{5\to 4}$, (c) mass of the top candidate, (d) mass of the Higgs candidate.
Each sample is weighted assuming an integrated luminosity of 1~ab$^{-1}$
with beam polarizations $(P_{e^-},P_{e^+})=(-0.8,+0.3)$.
In each of these 4 plots, all the event selection criteria
are applied except for the cut on the variable shown.
The arrows indicate the optimized cut values.
}
\label{fig:massfinal-6j}
\end{figure*}

We summarize the yields after applying each cut
for the case of polarized beams $(P_{e^-},P_{e^+})=(-0.8,+0.3)$
in Tab.~\ref{tab:allpol-6j},
where the yields are normalized assuming
an integrated luminosity of 1~ab$^{-1}$.
The resulting distributions for the thrust,
$Y_{5\to4}$, the top candidate mass, and the
Higgs candidate mass,
after applying all the other cuts,
are shown in Fig.~\ref{fig:massfinal-6j}.
The signal significance is estimated to be $3.7$, corresponding to the
measurement accuracy of the top Yukawa coupling of
$\left|\Delta g_t/g_t\right|=14\%$.
With unpolarized beams $(P_{e^-},P_{e^+})=(0.0,0.0)$,
the significance becomes $2.9$, corresponding to $\left|\Delta g_t/g_t\right|=17\%$.
\begin{table*}[hbt!p]
\caption{Summary of cuts in the analysis of the 6-jet + lepton mode, denoted as $6j$.
    We denote the 4-jet + 2-lepton mode as $4j$, and the 8-jet mode as $8j$.
Estimated yields are given assuming an integrated luminosity of 1~ab$^{-1}$
with beam polarizations $(P_{e^-},P_{e^+})=(-0.8,+0.3)$.
Refer to the text for the details of the $b$-tagging requirement and the mass cuts.}
\label{tab:allpol-6j}
\centering
\begin{ruledtabular}
\begin{tabular}{l......}
& \multicolumn{1}{c}{$t\overline{t}H(6j)$}
& \multicolumn{1}{c}{$t\overline{t}H(8j)$}
& \multicolumn{1}{c}{$t\overline{t}H(4j)$}
& \multicolumn{1}{c}{$tbW$}
& \multicolumn{1}{c}{$t\overline{t}Z$}
& \multicolumn{1}{c}{$t\overline{t}g^*\,(b\overline{b})$} \\
\hline
no cuts                & 282.3   &  289.5 &   68.3 &980738.5 & 2406.9 &1159.6 \\
single isolated lepton & 179.6   &   20.7 &   28.3 &340069.0 &  790.6 & 397.7 \\
thrust $<0.77$         & 145.7   &   18.5 &   19.2 &144999.0 &  616.7 & 266.0 \\
$Y_{5\to4} > 0.005$    & 125.5   &   16.6 &    9.2 & 12297.7 &  416.2 & 113.7 \\
$b$-tagging            &  49.0   &    1.3 &    2.9 &   172.9 &   53.3 &  37.8 \\
mass cuts              &  39.5   &    1.2 &    0.4 &    23.0 &   33.9 &  13.2 \\
\end{tabular}
\end{ruledtabular}
\end{table*}

\section{Analysis of the 8-jet mode}
\label{sec:8jet}
\subsection{Isolated lepton rejection}
In the 8-jet analysis, the 6-jet + lepton mode
can become a source of background as a result of
splitting the jets by the jet clustering procedure.
To reduce this kind of background, we look for isolated leptons using 
the same prescription used in the 6-jet + lepton analysis.
Events containing one or more isolated leptons are discarded in the 8-jet analysis.
This procedure ensures that the 6-jet + lepton and 8-jet samples are
statistically independent from each other,
allowing for a straightforward combination of the two results.

\subsection{Event shape}
Similarly to the 6-jet + lepton analysis, the thrust variable $T$
is used to reduce the $t\overline{t}$ background.
It is found that $T<0.7$ is optimal for the 8-jet mode.

\subsection{Jet clustering}
In the 8-jet mode analysis, the jet clustering is performed over
all particles in the event to form eight jets.
We keep the jet transition values $Y_{n\to n-1}$.
The value for $Y_{8\to 7}$ is found to be useful in discriminating
$t\overline{t}H$ events from $t\overline{t}$ events.
We require $Y_{8\to7}>0.0009$ in the event selection.

\subsection{Identification of heavy flavor jets}
We follow a similar procedure as in the 6-jet + lepton mode 
for the identification of $b$-jets.
We use a different optimization for the tight $b$-jet,
which is modified to be $(Q,N_Q)=(3.0,2)$.
The definition of the loose $b$-jet remains the same.

\subsection{Top and Higgs reconstruction}

The Higgs candidate ($H\to b\overline{b}$) is formed by
requiring one tight $b$-jet and one loose $b$-jet.
One of the top candidates is required to contain a
tight $b$-jet, while the other top is required to have
(at least) a loose $b$-jet.
Because there are multiple possible combinations of jets,
we define the quantity $\chi^2$
similarly to the 6-jet + lepton mode
as
\begin{widetext}
\begin{equation}
\chi^2 =
\left(\frac{m_{j_1j_2}-M_H}{\sigma_H}\right)^2
+\left(\frac{m_{j_3j_4j_5}-M_t}{\sigma_t}\right)^2
+\left(\frac{m_{j_6j_7j_8}-M_t}{\sigma_t}\right)^2
+\left(\frac{m_{j_3j_4}-M_W}{\sigma_W}\right)^2 
+\left(\frac{m_{j_6j_7}-M_W}{\sigma_W}\right)^2,
\label{eq:chi2-8j}
\end{equation}
\end{widetext}
and choose the combination of jets
which minimizes the $\chi^2$ value.
Here, $j_1$ and $j_2$ are used to form the Higgs candidate.
The three jets $j_3$, $j_4$, and $j_5$ are
used to reconstruct the first top candidate,
while $j_6$, $j_7$, and $j_8$ are used to reconstruct
the second top candidate.
The same values for $\sigma_t$, $\sigma_W$, and $\sigma_H$
are used as in the 6-jet + lepton analysis.

Final cuts are applied on the invariant mass of the top and Higgs candidate.
For both top candidates, the mass is required to be in the range of
140~GeV $< m_{jjj} < $ 215~GeV.
The Higgs candidate mass is required to be in the range of
80~GeV $< m_{jj} < $ 150~GeV.

\subsection{Results}
\begin{figure*}[hbt!p]
\centering
\includegraphics[width=0.45\linewidth]{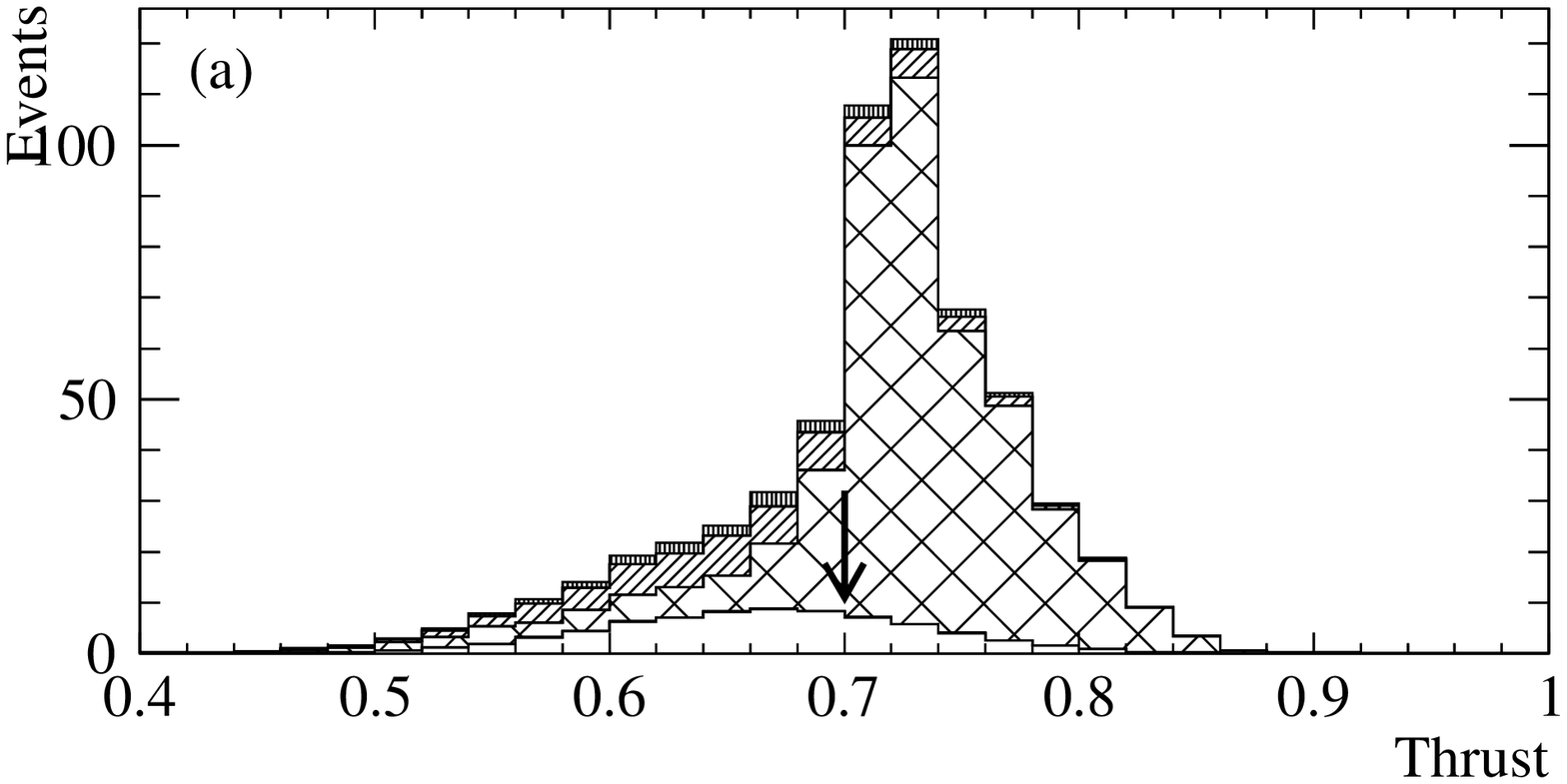}
\includegraphics[width=0.45\linewidth]{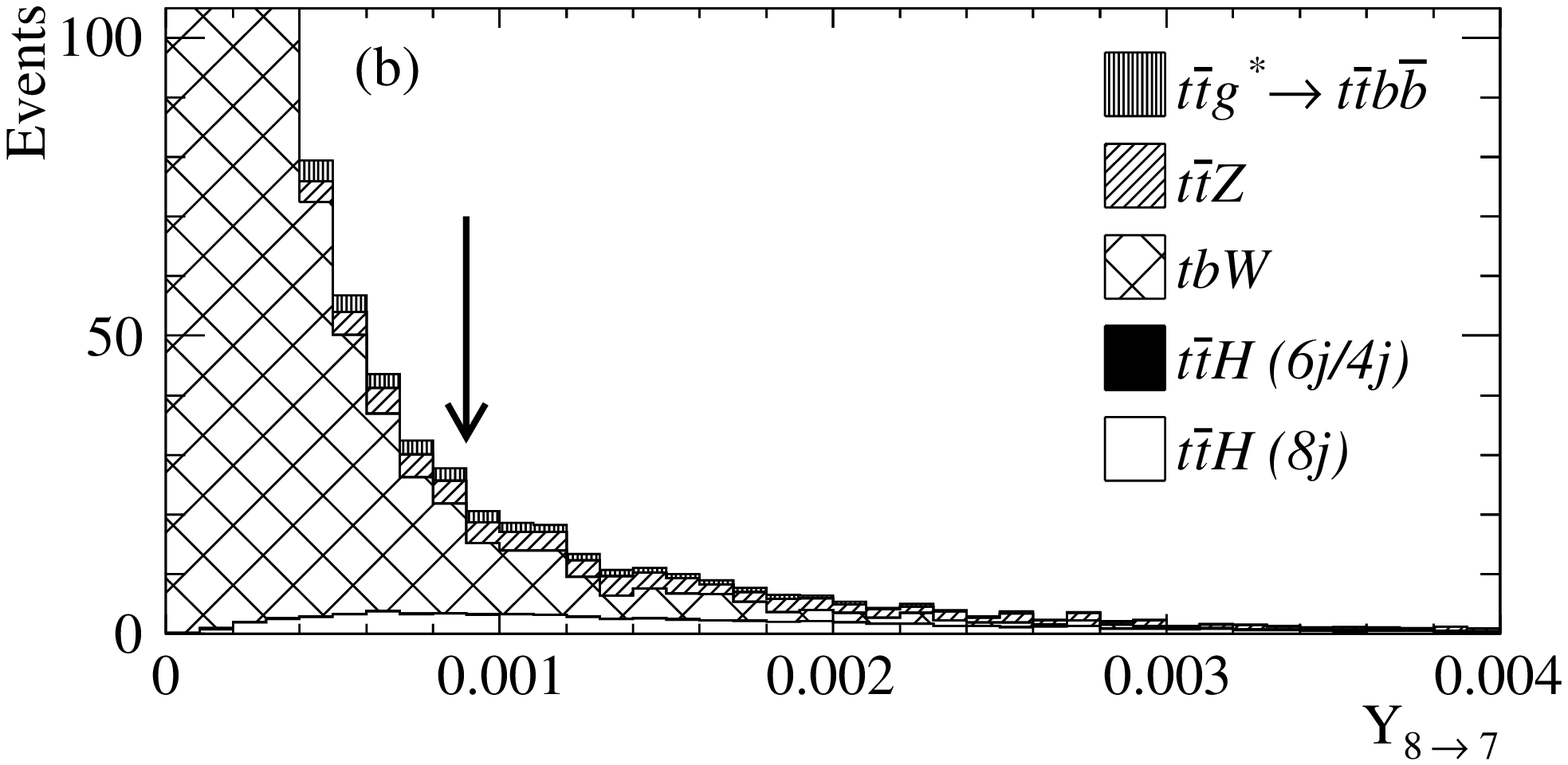}
\includegraphics[width=0.45\linewidth]{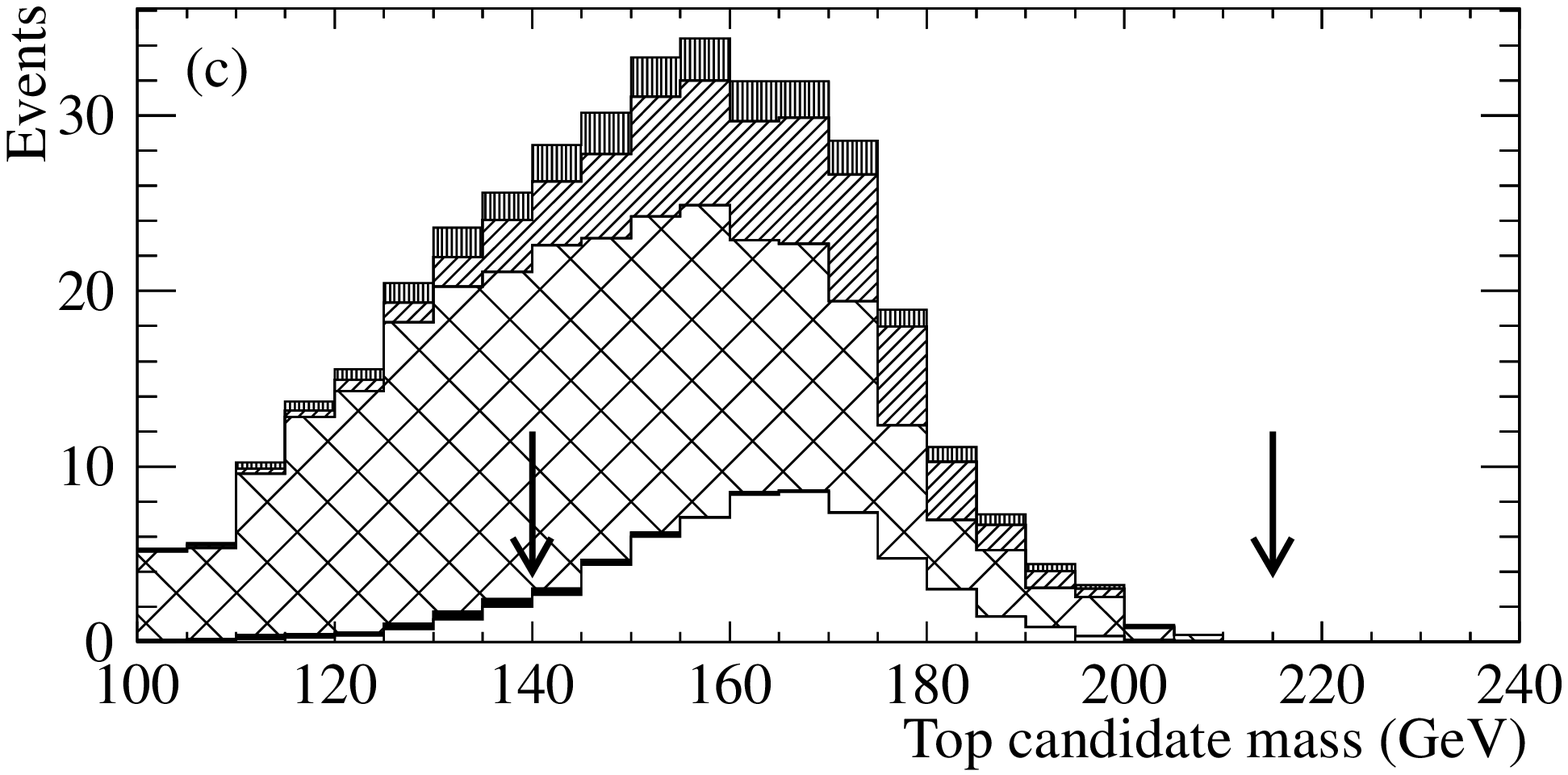}
\includegraphics[width=0.45\linewidth]{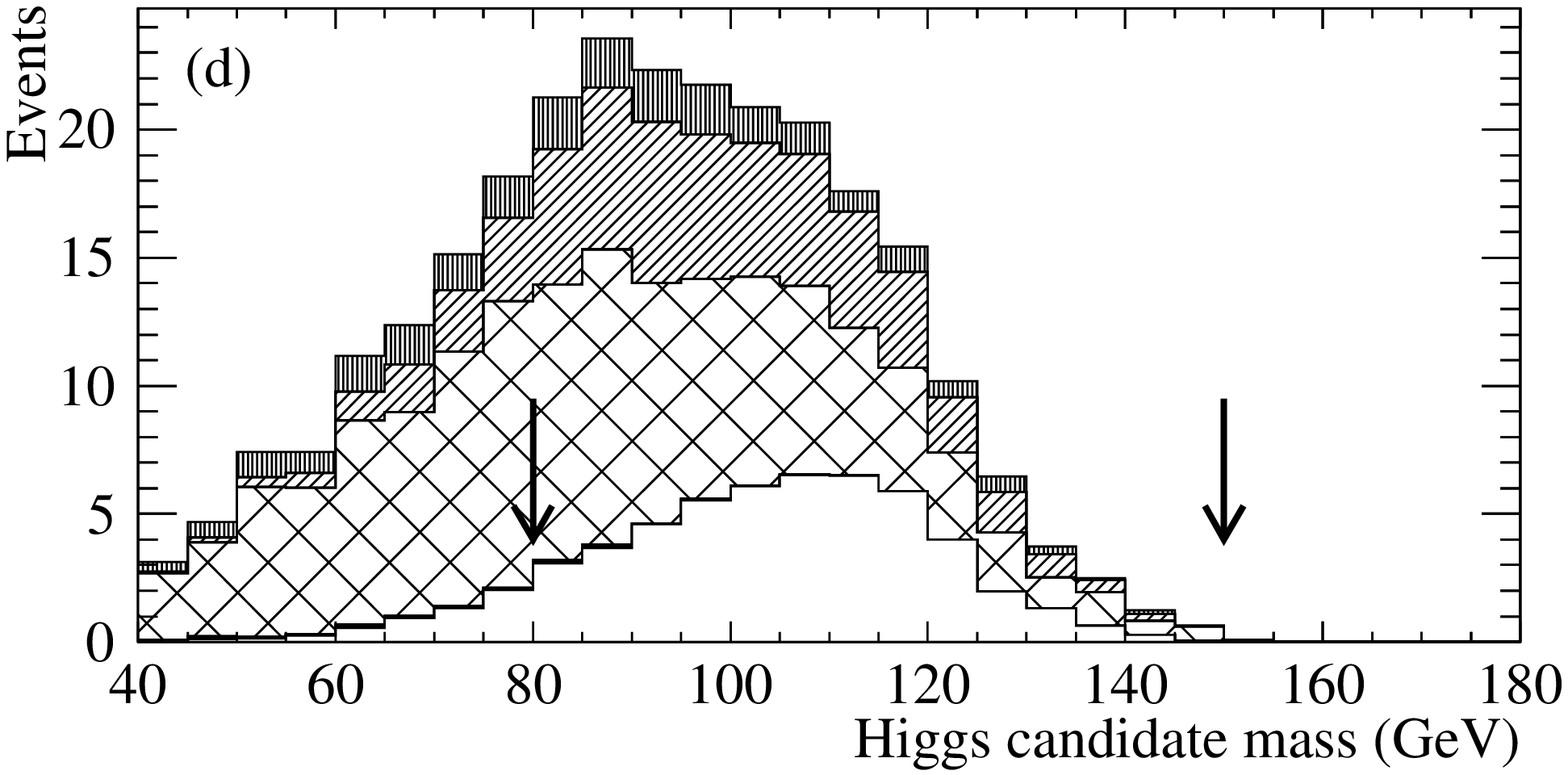}
\caption{
The distributions of the cut variables in the 8-jet analysis are shown:
(a) thrust, (b) $Y_{8\to 7}$, (c) mass of the top candidate, (d) mass of the Higgs candidate.
Each sample is weighted assuming an integrated luminosity of 1~ab$^{-1}$
with beam polarizations $(P_{e^-},P_{e^+})=(-0.8,+0.3)$.
In each of these 4 plots, all the event selection criteria
are applied except for the cut on the variable shown.
The arrows indicate the optimized cut values.
}
\label{fig:massfinal-8j}
\end{figure*}

The estimated signal yields are summarized in Tab.~\ref{tab:allpol-8j}
for the case of polarized beams $(P_{e^-},P_{e^+})=(-0.8,+0.3)$,
assuming an integrated luminosity of 1~ab$^{-1}$.
The resulting distributions for
the thrust, $Y_{8\to7}$, the top mass, and the Higgs mass
are shown in Fig.~\ref{fig:massfinal-8j}.
The signal significance in the 8-jet mode is 3.7,
corresponding to the measurement accuracy of the top Yukawa coupling of
$\left|\Delta g_t/g_t\right|=14\%$.
With unpolarized beams $(P_{e^-},P_{e^+})=(0.0,0.0)$,
the significance becomes 2.8, corresponding to
$\left|\Delta g_t/g_t\right|=18\%$.

\begin{table*}[hbt!p]
\caption{Summary of cuts in the analysis of the 8-jet mode, denoted as $8j$.
    We denote the 6-jet + lepton mode as $6j$, and the 4-jet + 2-lepton mode as $4j$.
Estimated yields are given assuming an integrated luminosity of 1~ab$^{-1}$
with beam polarizations $(P_{e^-},P_{e^+})=(-0.8,+0.3)$.
Refer to the text for the details of the $b$-tagging requirement and the mass cuts.}
\label{tab:allpol-8j}
\centering
\begin{ruledtabular}
\begin{tabular}{l......}
& \multicolumn{1}{c}{$t\overline{t}H$($8j$)}
& \multicolumn{1}{c}{$t\overline{t}H$($6j$)}
& \multicolumn{1}{c}{$t\overline{t}H$($4j$)}
& \multicolumn{1}{c}{$tbW$}
& \multicolumn{1}{c}{$t\overline{t}Z$}
& \multicolumn{1}{c}{$t\overline{t}g^*\,(b\overline{b})$} \\
\hline
no cuts                 & 289.5   &  282.3 &   68.3 &980738.5 & 2406.9 &1159.6 \\
reject isolated leptons & 266.3   &   85.6 &    6.6 &589716.0 & 1351.4 & 701.2 \\
thrust $<0.7$           & 167.7   &   44.0 &    2.7 &107227.0 &  818.0 & 311.5 \\
$Y_{8\to7} > 0.0009$    & 113.8   &   13.0 &    0.3 &  4048.1 &  349.6 &  67.1 \\
$b$-tagging             &  66.6   &    6.8 &    0.1 &   442.6 &   77.6 &  39.8 \\
mass cuts               &  50.1   &    0.4 &    0.0 &    75.6 &   47.6 &  14.1 \\
\end{tabular}
\end{ruledtabular}
\end{table*}

\section{Conclusions}
\label{sec:conclusions}
We have evaluated the accuracy of the top Yukawa coupling at
$\sqrt{s}=500$~GeV, taking into account the $t\overline{t}$ bound-state effects
for the $e^+e^-\to t\overline{t}H$ signal sample as well as the
$e^+e^-\to t\overline{t}Z$ background sample.
Other backgrounds considered were
$e^+e^-\to t\overline{b}W^-/\overline{t}bW^+\to bW^+\overline{b}W^-$ and
$e^+e^-\to t\overline{t}g^*\to bW^+\overline{b}W^-b\overline{b}$.
A simple cut-and-count analysis was performed for
the 6-jet + lepton and 8-jet signal decay modes.
We assume an integrated luminosity of 1~ab$^{-1}$.
Because the 6-jet + lepton sample and the 8-jet sample are statistically independent,
the combined significance can be computed simply by summing
the significances of the two modes in quadrature,
assuming Gaussian statistics.

With polarized beams $(P_{e^-},P_{e^+})=(-0.8,0.3)$,
the combined significance is 5.2,
corresponding to the measurement accuracy of the top Yukawa coupling of
$\left|\Delta g_t/g_t\right|=10\%$.
With unpolarized beams $(P_{e^-},P_{e^+})=(0.0,0.0)$,
the combined significance becomes 4.0,
corresponding to $\left|\Delta g_t/g_t\right|=13\%$.
Note that these numbers only take into account the statistical uncertainty.

The largest background contribution is the
$e^+e^- \to t\overline{t}Z$ process
which survives the event selection procedure
primarily because of the overlapping of the dijet mass
for the $Z$ and the Higgs.
This can be reduced by improving
the jet energy resolution and the
jet clustering procedures, which in turn
improves the mass resolution of the Higgs candidate.
The second largest background contribution is
the $e^+e^- \to tbW$ process.
Thus it will be critical to be able to model the
$e^+e^- \to tbW$ cross section accurately,
particularly in the tails of its kinematically allowed region.

Our results indicate that the measurement of the top Yukawa coupling
is possible down to the 10\% level of
statistical precision at the ILC with $\sqrt{s}=500$~GeV
after taking into account the $t\overline{t}$ bound-state effects,
which agrees with previous predictions~\cite{Juste:2006sv}.
It will be critical to reduce the systematic effects down to the level
comparable to the statistical uncertainties.
We expect the systematic uncertainties coming from the determination
of the background rates to be the dominant effect.
The amount of $tbW$ background
can be estimated by measuring the $tbW$ cross section
at $\sqrt{s}=500$~GeV.
The $t\overline{t}$ bound-state effects must also be verified
by measuring the $t\overline{t}$ cross section at its production threshold
($\sqrt{s}\approx 350$~GeV)
which will be used to estimate the rate of 
the $e^+e^-\to t\overline{t}H$ signal and the $e^+e^-\to t\overline{t}Z$ background.
For this, it will be necessary to measure the differential cross section
of $e^+e^-\to t\overline{t}$
in order to separate the 
Higgs-exchange contribution via the $t$-channel
which itself contains the top Yukawa coupling.

On the theoretical front,
it will be desirable to reduce the uncertainties
in the production cross section
coming from loop corrections,
which will be critical for precise background estimation.
For the $e^+e^-\to t\overline{t}$ process,
the electroweak corrections
are known at the 1-loop level~\cite{Beenakker:1991ca},
with further improvements expected in the coming years.
QCD corrections are already known at the 3-loop
level~\cite{Grunberg:1979ru,Jersak:1981sp,Chetyrkin:1996cf,Chetyrkin:1997mb,Chetyrkin:1997pn}.
For the $e^+e^-\to t\overline{t}Z$ process,
the known QCD corrections at the 1-loop level~\cite{Lei:2008jx}
include the $t\overline{t}$ bound-state effects.
Since our study also incorporates the $t\overline{t}$ bound-state effects,
it will be necessary to calculate the higher order corrections
in order to properly estimate the theoretical uncertainties
in the $e^+e^-\to t\overline{t}Z$ cross section.

\begin{acknowledgments}
The authors wish to express their gratitude
to all members of the ILC physics subgroup
\cite{jlc} for useful discussions. 
Among them, A.~Ishikawa and S.~Uozumi deserve special mention
for their work during the initial stages of this work.
The authors would like to thank J.~Kanzaki for discussions
on the discovery potential of the $t\overline{t}H$ process at the LHC.
This work is supported in part by the Creative Scientific Research Grant
No.~18GS0202 of the Japan Society for Promotion of Science (JSPS),
the JSPS Core University Program,
and the JSPS Grant-in-Aid for Scientific Research No.~22244031.
\end{acknowledgments}

\bibliography{tth-paper.bib}

\end{document}